\documentclass[aps,pra,twocolumn]{revtex4-2}
\usepackage[utf8]{inputenc}
\usepackage{bm}
\usepackage{bbm}
\usepackage{bbold}
\usepackage{amsfonts}
\usepackage{amsmath}
\usepackage{amssymb}
\usepackage{graphicx}
\usepackage{mathtools}
\usepackage{upgreek}
\usepackage{complexity}
\usepackage{xfrac}
\usepackage{soul}
\usepackage{xcolor}
\usepackage{dsfont}

\usepackage[colorlinks=true,citecolor=blue]{hyperref}
\hypersetup{colorlinks=true,citecolor=blue,linkcolor=blue,urlcolor=blue}
\usepackage{url}

\usepackage{txfonts}
\DeclareSymbolFont{matha}{OML}{txmi}{m}{it}
\DeclareMathSymbol{\varv}{\mathord}{matha}{118}

\begin{document}

\title{Ground state-based quantum feature maps}

\author{Chukwudubem Umeano}
\affiliation{Department of Physics and Astronomy, University of Exeter, Stocker Road, Exeter EX4 4QL, United Kingdom}

\author{Oleksandr Kyriienko}
\affiliation{Department of Physics and Astronomy, University of Exeter, Stocker Road, Exeter EX4 4QL, United Kingdom}

\date{\today}

\begin{abstract}
We introduce a quantum data embedding protocol based on the preparation of a ground state of a parameterized Hamiltonian. We analyze the corresponding quantum feature map, recasting it as an adiabatic state preparation procedure with Trotterized evolution. We compare the properties of underlying quantum models with ubiquitous Fourier-type quantum models, and show that ground state embeddings can be described effectively by a spectrum with degree that grows rapidly with the number of qubits, corresponding to a large model capacity. We observe that the spectrum contains massive frequency degeneracies, and the weighting coefficients for the modes are highly structured, thus limiting model expressivity. Our results provide a step towards understanding models based on quantum data, and contribute to fundamental knowledge needed for building efficient quantum machine learning (QML) protocols. As non-trivial embeddings are crucial for designing QML protocols that cannot be simulated classically, our findings guide the search for high-capacity quantum models that can largely outperform classical models.
\end{abstract}

\maketitle


\section{Introduction}
Quantum computing offers a powerful platform for data processing and artificial intelligence (AI) solutions. Quantum machine learning (QML) has been established as a field of research, where AI problems are solved with the help of quantum circuits \cite{Biamonte2017,Benedetti2019rev}. To date, QML has been applied to various machine learning problems and applications. These include classification \cite{Cong2019,SamuelChen2022,Monaco2023}, regression \cite{Mitarai_2018}, generative modelling \cite{Zoufal2019,kyriienko2022protocols,PaineQQM2023}, scientific computing \cite{KyriienkoPaine2021}, graph analysis \cite{Albrecht2023}, federated learning \cite{SamuelChen2021}, and many others. Here, quantum computers are suggested as potential linear algebra accelerators \cite{Biamonte2017}, or as a platform for building quantum models with improved expressivity or generalization properties \cite{Schuld2019feature,Caro2022}. The success of QML implementations depends on many ingredients, and there are key challenges to be resolved.

One of the challenges of quantum machine learning corresponds to loading data into quantum states. Here, the power of QML ultimately relies on the way models are built, and the choice of data embedding \cite{Schuld2019feature,SchuldSweke2021PRA,Caro2022,Goto2021PRL}. This is often done in the form of quantum feature maps --- circuits that map data points into quantum states, for instance via parameterized operations. Major examples include rotation-based quantum feature maps and amplitude-encoded data. However, the former are associated to Fourier models \cite{SchuldSweke2021PRA,schuld2021supervised} that may become tractable by other approaches \cite{Landman2022random}, while the latter are based on quantum RAM or access to oracles which are difficult to implement \cite{Biamonte2017,Giovannetti2008}. Recently, new approaches were put forward, including feature maps based on linear combinations of unitaries (LCUs) \cite{williams2023quantum,Childs2012,Childs_2017}. Finally, there is a significant part of research in QML that concerns quantum data \cite{schatzki2021entangled,Caro2022,Larocca2022PRXQ,Cong2019} --- access to quantum states (pure or mixed) that originate from quantum processes or low energy states of some quantum systems. Intrinsically, these data emerge from state preparation that depend on system's properties \cite{umeano2023learn}.

Recent works have shown great promise of analysing quantum data and learning from experiments \cite{Caro2022,Larocca2022PRXQ,Huang2022}. They have identified the generalization advantage and ability to learn from smaller numbers of samples. At the same time, the relation between QML models for learning on quantum and classical data remains missing, in particular regarding the data embedding step. The emerging understanding of QML is that quantum embedding can be the prime source of advantage \cite{Cerezo2023CSIM}, and is needed for designing classically intractable models with trainable variational ansatze. 

In this work, we present and formalize a protocol for embedding data $x$ into a ground state $|\psi_{\mathrm{G}}\rangle$ of a nontrivial Hamiltonian $\hat{\mathcal{H}}(x)$. As we prepare the ground state with some process, this can be seen as a quantum feature map based on a ground state preparation (GSP) protocol. We analyze properties of ground state-based feature maps. Specifically, we ask the question: can we represent models with GSP-based embeddings as Fourier-type quantum models? To answer this question, we establish a formal connection of GSP embedding with adiabatic state preparation. We find that ground state-based feature maps can be seen as models with frequency gap spectrum that grows at least as a high-degree polynomial in the number of qubits $N$, and can be exponential for complex Hamiltonians $\hat{\mathcal{H}}(x)$. We describe this as a QML model with high \emph{capacity}. Simultaneously, this spectrum features a large degeneracy of frequencies (scaling combinatorially), and highly structured coefficients for the emerging basis that lead to specific models and limited \emph{expressivity}. Our results indicate that QML models of similar performance can be built once multiple re-uploadings and structured circuits are used.
\begin{figure*}[t]
\centerline{\includegraphics[width=1.0\linewidth]{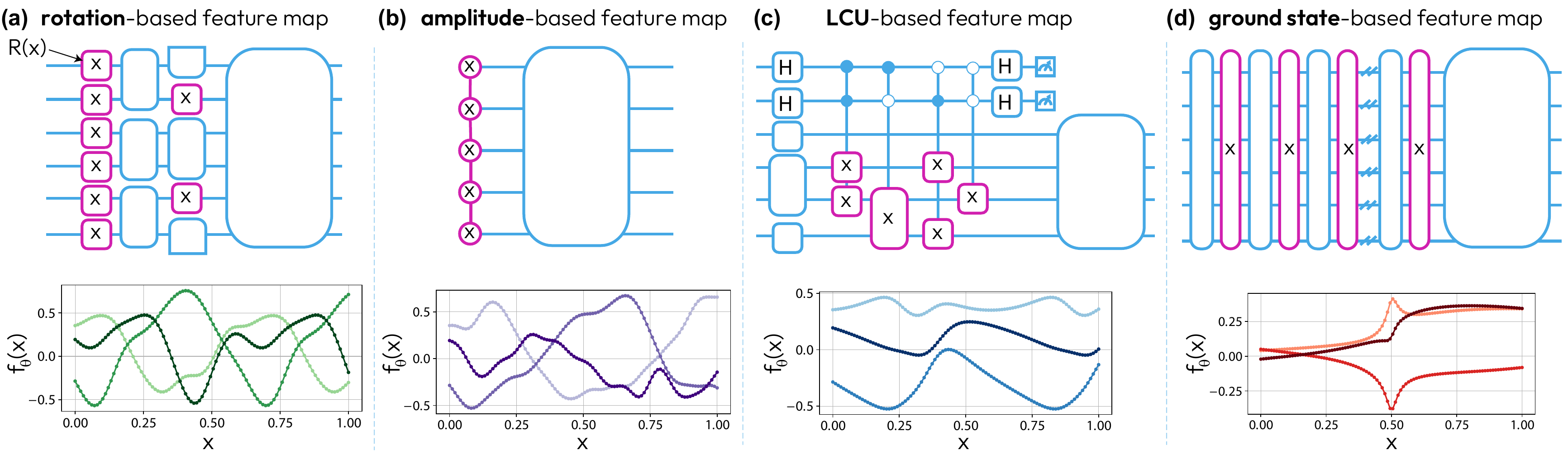}}
\caption{Different classes of quantum feature maps for data embedding. \textbf{(a)} Rotation-based feature maps as quantum circuits that lead to Fourier-type models. Here and hereafter, examples of  models are shown below each circuit. \textbf{(b)} Amplitude encoding, where $x$-dependence is embedded into (exponentially many) amplitudes of an input quantum state. \textbf{(c)} LCU-based feature maps provide non-Fourier spectrum due to measurement. \textbf{(d)} Ground state feature maps embed data into very specific states, and can be considered as quasi-Fourier models with exponentially scaling of the spectrum (discussed in this work).}
\label{embeddings}
\end{figure*}


\section{Model}
A feature map is the operator $\hat{\mathcal{U}}_{\varphi}(x)$ which embeds the classical data $x$ into a state in the Hilbert space. This corresponds to the map $x \mapsto |\psi(x)\rangle=\hat{\mathcal{U}}_{\varphi}(x)|\psi_0\rangle$, where $|\psi_0\rangle$ is some trivial initial state \cite{Mitarai_2018,Schuld2019feature,Goto2021PRL}. A standard choice is the rotation-based embedding, made up of single-qubit Pauli rotations, 
\begin{equation}\label{eq:rot_embed}
    |\psi(x)\rangle = \prod_{i=1}^N \hat{R}^i_{\mathrm{\alpha}}(\phi_i(x))|\psi_0\rangle,
\end{equation}
with $\alpha \in \{\mathrm{X},\mathrm{Y},\mathrm{Z}\}$. If we set $\phi_i(x)=\phi(x)=\eta x$, where $\eta$ is some constant, this encoding naturally leads to a Fourier-type model with a simple frequency spectrum (see \cite{SchuldSweke2021PRA} and also discussion later in the text). If we wish to extend the frequency spectrum, we can add more $x$-dependent rotations to extended the spectrum, in the process known as data re-uploading \cite{PerezSalinas2020datareuploading}. We visualize an example of a rotation-based QML circuit in Fig.~\ref{embeddings}(a), also demonstrating below a few model instances that show harmonic behavior. We note that rotation-based quantum feature maps are widespread in the near-term QML pipeline, as they do not require many resources (deep circuits), typically have good generalization \cite{Caro2022} and their model capacity (number of available frequencies) can be readily controlled. However, these can be prone to classical simulability in the absence of deep variational circuits \cite{Cerezo2023CSIM}.

Another approach to embed data corresponds to an amplitude-based encoding \cite{Biamonte2017}, where one directly encodes the inputs into the $2^N$ different amplitudes of the quantum state in the computational basis, $|\psi(x)\rangle = \sum_{j=1}^{2^N} c_j(x)|j\rangle$, 
%
%
where $\{|j\rangle\}$ are the computational basis elements, and $\sum_j c_j(x)^2 = 1$. The amplitude-based encoding is more suitable to large-scale fault-tolerant quantum computers due to the resources needed to implement it. Depending on the structure of the set of amplitudes, this data encoding method requires either arbitrary state preparation strategies or qRAM \cite{Biamonte2017,Giovannetti2008}, both non-trivial protocols. The $x$-dependence is now found in exponentially many amplitudes of the state. One can also connect amplitude-based QML with rotation-based models by introducing a phase feature map and utilizing the quantum Fourier transform \cite{kyriienko2022protocols}. Then the rotation-based map \eqref{eq:rot_embed} follows $\phi_j(x)=2\pi x/2^j$ scaling, and leads to exponentially many frequencies that are non-degenerate. We visualize a circuit with an amplitude-based feature map in Fig.~\ref{embeddings}(b).

A qualitatively distinct way to embed data relies on effectively non-unitary maps \cite{williams2023quantum}. Namely, we may encode our inputs using a non-unitary operator $\hat{\mathcal{A}}_{\varphi}(x)$ rather than the unitary circuit $\hat{\mathcal{U}}_{\varphi}(x)$ which is typically used. We can decompose any operator as a linear combination of unitaries (LCU), $\hat{\mathcal{A}}_{\varphi}(x)=\sum_k\alpha_k(x)\hat{U}_k(x)$, encoding this operator into an extended Hilbert space using ancilla qubits \cite{Childs2012,Childs_2017}. This type of embedding protocol is useful when we desire to generate a quantum model with non-Fourier modes. For example, an LCU circuit is used for implementing an orthogonal Chebyshev feature map, which encodes Chebyshev polynomials of exponentially growing degree into the computational basis state amplitudes \cite{williams2023quantum}. We visualize an example of the LCU-based quantum feature map in Fig.~\ref{embeddings}(c), and show several model instances below, which demonstrate non-harmonic behavior coming from $x$-dependent normalization.

Finally, in this study we formalize another data encoding method considering the ground state preparation process on quantum computers. Here, the feature $x$ parameterizes a quantum Hamiltonian $\hat{\mathcal{H}}(x)$, and the mapping process assumes preparing its ground state $|\psi_{\mathrm{G}}(x)\rangle$. This can be achieved by several procedures, including either non-unitary (cooling) or unitary state preparation, $|\psi_{\mathrm{G}}(x)\rangle = \hat{\mathcal{U}}_{\varphi}(x)|\psi_0\rangle$, with a map $\hat{\mathcal{U}}_{\varphi}(x)$. We note that this process is often assumed when working with so-called quantum data \cite{schatzki2021entangled,Caro2022,Larocca2022PRXQ,Cong2019}, and has recently been connected to a hidden feature map process \cite{umeano2023learn}. However, there is an open question: how can one prepare the ground state-based datasets, and how does the corresponding QML model compare to other approaches? We visualize an exemplary quantum circuit for the GSP-based feature map in Fig.~\ref{embeddings}(d), elucidating details in the following, and show the related models below the circuit.


\subsection*{Ground state preparation and feature maps}\label{GSP}
To analyze formally the process of data embedding into ground states of parametrized Hamiltonians, we need to establish a suitable framework. Here, we discuss possible options and motivate the choice of adiabatic quantum state preparation as a bridge to QML based on Fourier-type models that is universally adopted.

We begin by recalling the different methods for GSP. One choice is to employ variational approaches, using either the variational quantum eigensolver (VQE) or the quantum approximate optimization algorithm (QAOA) \cite{Tilly_2022,farhi2014quantum,Pagano_2020}. If we wish to prepare ground states for a set of feature values, we have to train the variational circuit for each input value separately. This means that the $x$-dependence is implicitly hidden within the optimized parameters of the ansatz, $|\psi_{\mathrm{G}}(x)\rangle = \hat{\mathcal{U}}(\theta_{\mathrm{opt}}(x))|\psi_0\rangle$, and the functional dependence of the optimal angles $\theta_{\mathrm{opt}}(x)$ is generally non-trivial (and potentially has non-analytic jumps). This means that QML models built on top of ground states prepared in this fashion do not pertain well to analysis as a Fourier model. Next, GSP protocols based on effective thermalization or imaginary time evolution are of the LCU type, and thus cannot be connected to Fourier-type models. For the unitary GSP options, one possibility is to postulate that there exists a fixed circuit that connects an input state to the ground state. We analyze this approach in the Discussion section and the Appendix. Finally, one of the widespread ground state preparation methods that can help us to unravel the properties of GSP embedding is the ground state preparation procedure based on an adiabatic quantum protocol. This is the framework we concentrate on hereafter.

\subsection*{Adiabatic ground state preparation as a feature map}\label{AdiabaticGSP}

Let us proceed by describing the adiabatic quantum ground state preparation. Here we follow the standard description from Albash and Lidar \cite{AlbashLidarRMP2018}. The adiabatic GSP represents an annealing-type strategy, where a Hamiltonian is gradually tuned in time from a simple one ($\hat{\mathcal{H}}_0$) to a difficult one (target Hamiltonian $\hat{\mathcal{H}}_1$). Specifically, we consider a feature-dependent target Hamiltonian that embeds a scalar parameter $x$, and the full time-dependent operator reads
\begin{align}
\label{eq:H(t;x)}
\hat{\mathcal{H}}(t;x) = (1- t/T) \hat{\mathcal{H}}_0 + (t/T) \hat{\mathcal{H}}_1(x),
\end{align}
where $T$ is a total evolution time. The corresponding quantum propagator can be written as
\begin{align}
\label{eq:UT(x)}
\hat{\mathcal{U}}_T(x) = \mathcal{T}\left\{ \exp\left[\int\limits_{0}^{T} dt \hat{\mathcal{H}}(t;x)\right] \right\},
\end{align}
where $\mathcal{T}\{\cdot\}$ denotes the time-ordering operator and effectively represents a ground state-based $x$-feature map. Acting on the suitable initial state $|\psi_0\rangle$, that is, the ground state of $\hat{\mathcal{H}}_0$, we prepare the ground state of $\hat{\mathcal{H}}_1(x)$ as $|\psi_\mathrm{G}(x)\rangle = \hat{\mathcal{U}}_T(x)|\psi_0\rangle$, assuming that adiabaticity conditions are met (discussed later). Intrinsically $\hat{\mathcal{U}}_T(x)$ is a non-Fourier map, as it contains generally non-commuting parts $\hat{\mathcal{H}}_{0,1}$, such that the diagonalization becomes basis dependent (discussed in \cite{KyriienkoElfving2021}). As our goal is to analyze its properties while comparing to previously known models, we recast $\hat{\mathcal{U}}_T(x)$ into the form that admits an approximate Fourier-type representation. To break up the evolution with the non-commuting target and easy Hamiltonians, we employ a Trotter decomposition of the adiabatic state preparation, as is routinely done when implementing it digitally \cite{Barends2016,AlbashLidarRMP2018}. The corresponding feature map \eqref{eq:UT(x)} is decomposed into $M$ short steps,
\begin{align}
\label{eq:UT(x)_digital}
\hat{\mathcal{U}}_T(x) \approx \prod_{m=1}^{M} \left[ e^{-i(\Delta t^2 m /T) \hat{\mathcal{H}}_1(x) } \cdot e^{-i\Delta t \left(1 - m \Delta t/T\right) \hat{\mathcal{H}}_0 } \right],
\end{align}
where the time step $\Delta t = T/M$. The number of required steps for the high-fidelity GSP scales as $M = \mathcal{O}(T^2 \mathrm{poly}(N))$ \cite{AlbashLidarRMP2018}. We can rewrite the digital evolution operator in the form where $x$-dependent operators and other basis transforms are separated, recasting \eqref{eq:UT(x)_digital} as $\hat{\mathcal{U}}_T(x) \approx \prod_{m=1}^{M} \left[ \exp\{-i t_m x \hat{H}_\mathrm{G} \} \cdot \hat{W}_m \right]$. Here we introduce step-dependent operators $\hat{W}_m$ absorbing $x$-independent evolution, set $t_m = \Delta t^2m/T$, and assign a generator for the $x$-dependent unitary as $\hat{H}_\mathrm{G}$.

\subsection*{Feature map spectrum}\label{Spectrum}

To analyze formally the properties of the ground state-based feature map, we use the spectral decomposition
\begin{align}
\label{eq:spectral}
\exp\{-i t_m x \hat{H}_\mathrm{G} \} = \hat{V} \left( \sum_{\ell=1}^{|\Lambda|} \exp(-i t_m \lambda_\ell x) \hat{P}_{\ell} \right) \hat{V}^\dagger,
\end{align}
where $\hat{H}_\mathrm{G}$ is diagonalized by the basis transformation $\hat{V}$, and $\Lambda = \{ \lambda_\ell\}_{\ell}$ is a set of unique (non-repeated) eigenvalues with cardinality $|\Lambda|$. We introduce the corresponding subspace projectors $\{\hat{P}_\ell\}_\ell$ that collect eigenstate projectors for the same eigenvalue $\lambda_\ell$ (if it is degenerate), noting that $\sum_{\ell=1}^{|\Lambda|} \hat{P}_\ell = \mathbb{1}$. Then, similar to the procedure used in Schuld et al. \cite{SchuldSweke2021PRA}, the products of operators in $\hat{\mathcal{U}}_T(x)$ can be recast as $\hat{W}_{m+1} \exp\{-i t_m x \hat{H}_\mathrm{G} \}  \hat{W}_m = \sum_{\ell=1}^{|\Lambda|} \exp(-i t_m \lambda_\ell x) \hat{w}_{\ell}^{(m)}$, with $\hat{w}_{\ell}^{(m)} = \hat{W}_{m+1} \hat{V}^\dagger \hat{P}_{\ell} \hat{V} \hat{W}_m$ being the transition matrix that performs a basis change at $m \rightarrow m+1$ step, for the $\ell$-th eigensector. The full product in $\hat{\mathcal{U}}_T(x)$ then involves different combinations of $x$-dependent exponents, generating a \emph{mode} spectrum for the embedding, which we denote as $\Sigma$. This is a set of frequencies formed by different combinations of elements in the set of sets $(\Delta t^2/T) \left\{ \{ m \lambda_\ell\}_{\ell=1}^{|\Lambda|} \right\}_{m=1}^M$. The largest element of $\Sigma$ is called the \emph{degree} of the spectrum \cite{SchuldSweke2021PRA,KyriienkoElfving2021}, which in this case corresponds to the largest eigenvalue of $\Lambda$ multiplied by $\sum_{m=1}^M m$. This leads to the mode spectrum degree $D_\Sigma = \max(\Lambda) \frac{M(M+1)}{2} \frac{\Delta t^2}{T}$. Similarly, the \emph{width} of the spectrum is defined as the number of distinct frequencies, denoted as $K_\Sigma \coloneqq (|\Sigma|-1)/2$. 
Next, we note that quantum models based on the embedding $\hat{\mathcal{U}}_T(x)$ are typically built as expectation values over the feature state,
\begin{align}
\label{eq:f_model}
f_{\theta}(x) = \langle \psi_{\mathrm{G}}(x)| \hat{\mathcal{M}}_{\theta} | \psi_{\mathrm{G}}(x) \rangle,
\end{align}
where $\hat{\mathcal{M}}_{\theta}$ is a measurement operator (can be adjusted variationally). As in \eqref{eq:f_model} the embedding appears twice, the corresponding model $f_{\theta}(x) = \sum_{\omega_{k,k'} \in \Sigma} c_{k,k'} \exp[-i(\omega_k - \omega_{k'})x]$ involves frequency differences --- gaps --- with coefficients $c_{k,k'}$ being products of associated basis changes contracted with the input state and the measurement operator. The relevant gap spectrum for the model is $\Omega = \{ \omega_k - \omega_{k'}: \omega_{k,k'} \in \Sigma \}$. The degree for the GSP gap spectrum is $D_\Omega = \max(\Lambda) M(M+1) \Delta t^2/T$. We remind that given that the number of steps scales as $M \propto T^2 \mathrm{poly}(N)$, this leads to the spectrum degree $D_\Omega \propto \max(\Lambda) T$, i.e. proportional to the annealing time $T$. To meet the adiabaticity conditions one requires $T = \tilde{\mathcal{O}}\left( 1/\Delta_{\min}^2 \right)$, where $\Delta_{\min}$ is the minimal annealing gap of $\hat{\mathcal{H}}(t;x)$ (over the interval from $t=0$ to $T$) defined as a distance between instantaneous first excited and ground states. This has tremendous consequences for the ground state-based feature maps, namely because of the fast growth of $T$ with the system size. We expand on this point with some GSP embedding calculations in the next section.


\section{Results}

\begin{figure}[t!]
\centerline{\includegraphics[width=1.0\linewidth]{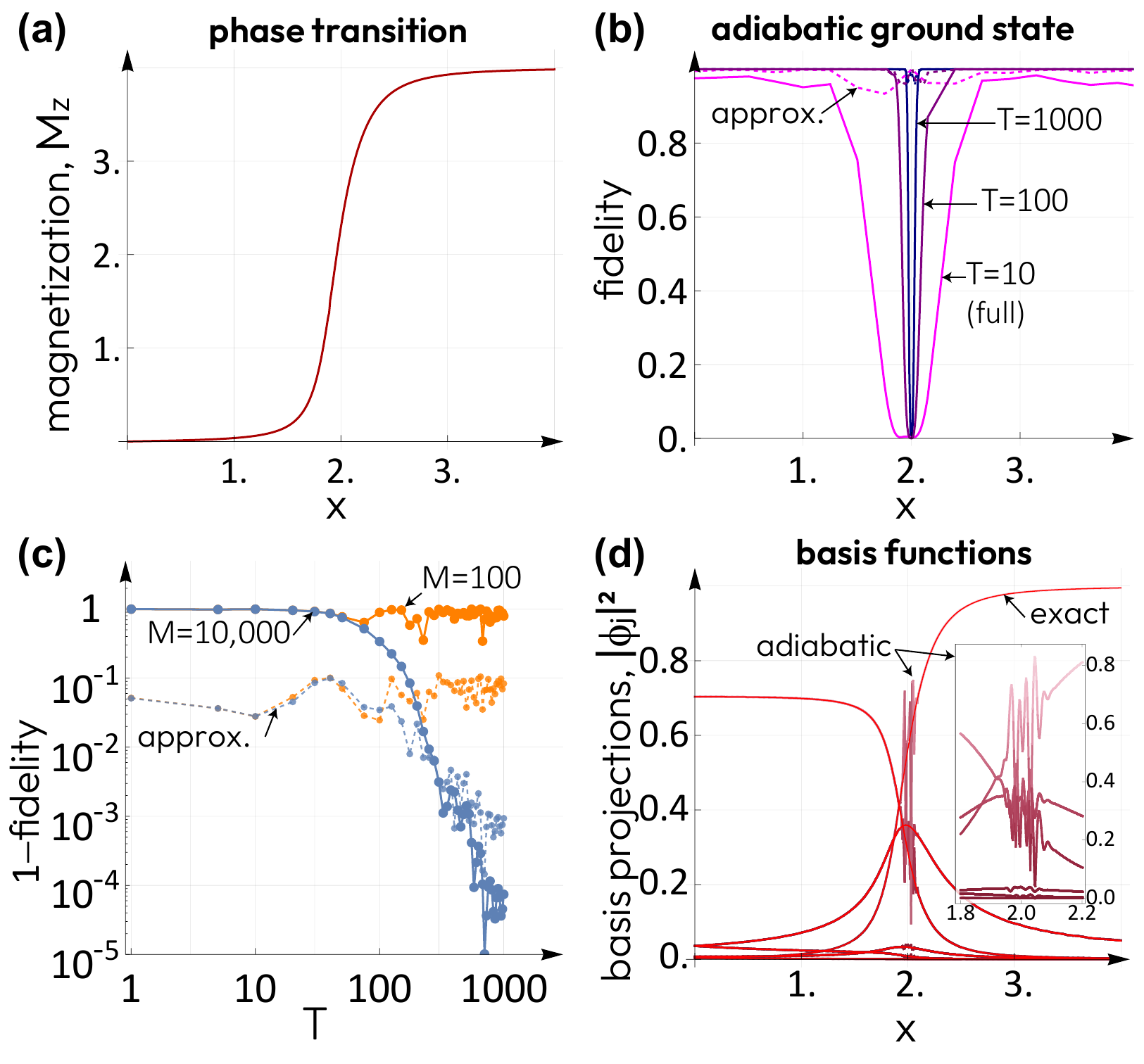}}
\caption{Analysis of the adiabatic ground state preparation for the Ising model with longitudinal field, and associated basis quality. \textbf{(a)} Expectation value of the magnetization operator for the exact ground states of the Ising model with longitudinal field (absolute value). \textbf{(b)} Fidelity of the adiabatic ground state as a function of $x$, plotted for different total evolution times $T$. Quality of the ground state deteriorates rapidly around the critical point, but the range of $x$ at which the full fidelity remains high increases with $T$. \textbf{(c)} Infidelity of the adiabatic ground state as a function of $T$. Only with sufficient Trotter steps can we converge to an accurate ground state at late times. For \textbf{(b)-(c)}, the solid curves show the full (in)fidelity while the dashed curves depict the approximate (in)fidelity. \textbf{(d)} Exact vs adiabatic basis functions. The errors introduced via Trotterization lead to highly oscillatoric basis functions for $x$ around the critical point.}
\label{adiabatic}
\end{figure}

We proceed by considering a concrete example and show the practical aspects of GSP-based feature mapping, as well as visualising the scaling. For this, we have chosen to work with the Ising-type Hamiltonian for a chain of $N$ qubits with periodic boundary conditions. The data-dependent Hamiltonian reads
\begin{align}
\label{eq:Hx_Ising}
\hat{\mathcal{H}}(x) = \sum\limits_{j=1}^{N} \hat{Z}_{j} \hat{Z}_{j+1} + x \sum\limits_{j=1}^{N} \hat{Z}_{j} + h \sum\limits_{j=1}^{N} \hat{X}_{j},
\end{align}
where $x$ is the parameter (feature) that we embed into the longitudinal part of the Hamiltonian, $\hat{H}_{\mathrm{Z}} \coloneqq \sum\limits_{j=1}^{N} \hat{Z}_{j}$. Here, we assume $\hat{Z}_{N+1} \equiv \hat{Z}_{1}$ (periodic boundary), and note that the transverse part of the field described by $\hat{H}_{\mathrm{X}} \coloneqq \sum\limits_{j=1}^{N} \hat{X}_{j}$ is introduced with $h < 1$, such that the basis deviates from the computational one. The interacting Ising-type Hamiltonian $\hat{H}_{\mathrm{ZZ}} \coloneqq \sum\limits_{j=1}^{N} \hat{Z}_{j} \hat{Z}_{j+1}$ is of the antiferromagnetic type. Here, we set the prefactor in front to be unity, implying that units of Ising interaction are used everywhere for measuring energy, and time is measured as inverse interaction strength set to one for convenience. The model features a phase transition as a function of $x$, where ground state ordering changes from the antiferromagnetic one to the staggered ferromagnetic. In Fig.~\ref{adiabatic}(a) we plot the absolute value of magnetization in Z direction over the ideal ground state $|\psi_{\mathrm{G}}^{\mathrm{(i)}}(x)\rangle$ of the Hamiltonian \eqref{eq:Hx_Ising}. In practice, this is obtained by the imaginary time evolution procedure $|\psi_{\mathrm{G}}^{\mathrm{(i)}}(x)\rangle \sim \exp[-\tau \hat{\mathcal{H}}(x)]|\psi_0\rangle$ with consequent state normalization. Here and hereafter we consider $h=0.2$ and $N=4$, such that a small scale system can be readily analyzed and visualized. The input state is chosen as $|\psi_0\rangle = |+\rangle^{\otimes N}$ being an equal superposition of all computational states, and we use $\tau = 20$.

\subsection*{Ising ground state preparation with Trotterized adiabatic evolution}

Next, we proceed to consider the unitary state preparation. We construct the adiabatic Hamiltonian $\hat{\mathcal{H}}(t;x)$ using $\hat{\mathcal{H}}_1(x)$ from Eq.~\eqref{eq:Hx_Ising}, and setting the initial Hamiltonian and state as $\hat{\mathcal{H}}_0 = -\hat{H}_X$ and $|\psi_0\rangle = |+\rangle^{\otimes N}$. We simulate the Trotterized unitary evolution \eqref{eq:UT(x)_digital} considering different annealing times $T$ and number of steps $M$. We perform ground state preparation for a full range of $x$ parameters, and test the fidelity for the GSP procedure. Specifically, we adopt two types of fidelity. First, we check a full overlap of the adiabatically prepared state $|\psi_{\mathrm{G}}(x;T,M)\rangle$ with the ideal state, $F(x) = |\langle \psi_{\mathrm{G}}^{\mathrm{(i)}}(x)|\psi_{\mathrm{G}}(x;T,M)\rangle|^2$. Note that this metric is demanding, as phases of individual components must match to yield unity fidelity for all $x$. Second, we also test the difference between computational state occupations, where the approximate fidelity is $\widetilde{F}(x) = 1 - \sum_{j=1}^{2^N} ||\langle j |\psi_{\mathrm{G}}(x;T,M)\rangle|^2 - |\langle j |\psi_{\mathrm{G}}^{\mathrm{(i)}}\rangle|^2|/2^N$, where probability vectors are compared.

In Fig.~\ref{adiabatic}(b) we show fidelities for several annealing schedules. First, when setting $T=10$ and $M=100$ we observe that the full fidelity remains very close to 1 outside of the critical region, but experiences a large drop in the window of $x$ surrounding the critical point $x_{\mathrm{cr}} = 2$ (solid magenta curve in Fig.~\ref{adiabatic}b). The corresponding approximate fidelity remains high throughout, but visibly deviates from unity (dashed magenta curve). Increasing the annealing time to $T = 100$ and $M = 1000$ we observe that fidelities are largely improved (purple curves in Fig.~\ref{adiabatic}b), but the full fidelity remains poor in the narrow critical window. This behavior continues even for $T=1000$ and $M=10,000$, albeit in the narrow window around $x_{\mathrm{cr}}$ (blue solid curve, Fig.~\ref{adiabatic}b).
\begin{figure}[t]
\centerline{\includegraphics[width=1.0\linewidth]{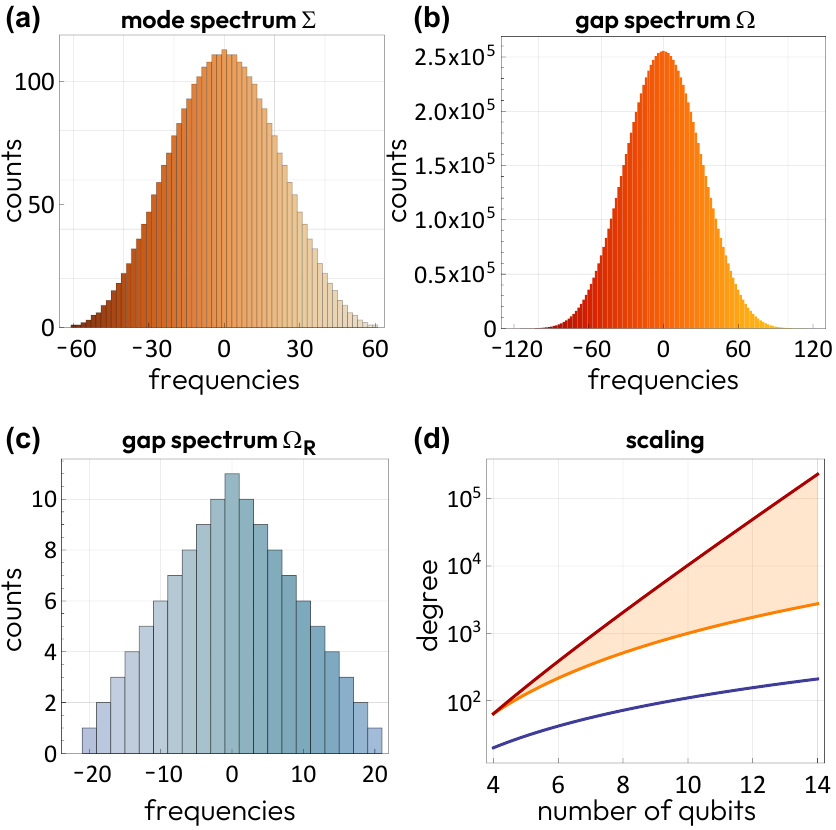}}
\caption{Spectrum of the ground state feature map. \textbf{(a)} Mode spectrum $\Sigma$ for the Ising-type ground state embedding, showing available frequencies (in the units of $\Delta t^2/T$) with corresponding counts of degenerate modes. Here we set $M=5$. \textbf{(b)} Gap spectrum $\Omega$ for the same Ising-type ground state embedding. \textbf{(c)} Gap spectrum for the rotation-based embedding $\Omega_\mathrm{R}$ based on the tower-type feature map of $N$ single-qubit rotations where prefactors of $x$ increase linearly with the qubit index. \textbf{(d)} Scaling of the gap spectral degree with the system size $N$, showing $\propto N^2$ scaling for the rotational embedding (blue solid curve), and  $\propto N^3$-to-$N 2^N$ scaling for GSP-type embedding, assuming different annealing gap scaling.}
\label{spectrum}
\end{figure}

We select a feature value of $x=1.9$ and plot the infidelity (being $1-F$) as a function of $T$. The results are shown in Fig.~\ref{adiabatic}(c). Blue curves and points show that infidelity drops to very low values as we make $T$ large (and significantly larger than $N$, assuming fine discretization with $M=10,000$). Similar GSP with the number of steps reduced to $M=100$ shows that infidelity remains very large, and the prepared state deviates from the ideal one even at $x=1.9$ (Fig.~\ref{adiabatic}c, yellow solid curve and dots). This confirms the analytical prediction that $M \propto T^2$ is needed for the mapping to work.

Finally, we test the consequences of imperfect state preparation on quantum machine learning models and their basis sets. In Fig.~\ref{adiabatic}(d) we show projections of the prepared state on the computational basis, $\phi_j (x)\equiv |\langle j |\psi_{\mathrm{G}}(x;T,M)\rangle|^2$, setting $T=1000$ and $M=10,000$ for the high fidelity GSP. This is also overlayed with ideal ground state projections shown as red curves in Fig.~\ref{adiabatic}(d). We reveal that the corresponding adiabatic GSP basis is effectively smooth, but at the critical region the functions experience rapid oscillations (see inset in Fig.~\ref{adiabatic}d). This shows that intrinsically the QML model for the adiabatic GSP works with high frequency components, but structures the coefficients in such a way that oscillations are suppressed, and largely degenerate low frequency modes are predominantly exploited. This also confirms the discussion of the spectral properties of ground state-based feature maps.

\subsection*{Spectrum of the Ising ground state feature map}

We proceed to study and visualize the underlying spectrum of QML models, now for the concrete example of the Ising-type embedding given by $\hat{\mathcal{H}}(x)$. Here, the generator corresponds to $\hat{H}_{\mathrm{G}}= \hat{H}_{\mathrm{Z}}$, and other $x$-independent parts of $\hat{\mathcal{H}}(x)$ are absorbed into basis changes upon Trotterization. There are $|\Lambda|=N+1$ unique eigenvalues of $\hat{H}_{\mathrm{G}}$ corresponding to $\Lambda = \{ \lambda_{\ell}=-N+2(\ell -1)\}_{\ell=1}^{N+1}$, starting from a non-degenerate polarized state and increasing degeneracy peaking at zero eigenvalue \cite{KozinKyriienko2019}. Based on the maximal eigenvalue for the chosen $\hat{H}_{\mathrm{G}}$ being $\max(\Lambda) = N$, we observe that the degree for the mode spectrum grows as $D_{\Sigma} = N T M(M+1)/2 M^2$ and $D_{\Omega} = N T M(M+1)/M^2$. The associated width of these spectra are equal to half-the-degree, $K_{\Sigma,\Omega} = D_{\Sigma,\Omega}/2$, as a consequence of the generator's spectrum. 

We proceed to visualize the shape of the GSP feature map spectrum in Fig.~\ref{spectrum}. The histogram of frequencies for the mode spectrum is presented in Fig.~\ref{spectrum}(a). Here we have assumed a modest number of Trotter steps ($M=5$) to aid the analysis, and plot frequency values in units of $\Delta t^2/T$. We observe the characteristic shape of normally distributed frequencies with a notable degeneracy. As we go from the mode spectrum to the gap spectrum, the shape remains the same, while the degeneracy grows to large numbers (Fig.~\ref{spectrum}b). This behavior is only more pronounced for larger $M$ required for GSP, and underlines a huge QML model capacity based on ground state embeddings.

To put these results into perspective, we compare GSP-based spectra with the standard rotation-based feature map. We consider a feature map of the type shown in Eq.~\eqref{eq:rot_embed}, where $N$ rotations have an increasing prefactor corresponding to the tower feature map \cite{KyriienkoPaine2021}. In this case we have $|\Lambda_{\mathrm{R}}| = N(N+1)/2 + 1$ unique eigenvalues, and in the absence of re-uploading the corresponding mode spectrum is flat (frequencies are non-degenerate). The gap spectrum $\Omega_{\mathrm{R}}$ for the rotation-based feature map is shown in Fig.~\ref{spectrum}(c), and features a linear decay of counts from zero frequency to the maximal frequency (degree) of $D_{\Omega_{\mathrm{R}}}=N(N+1)$, as well as showing an overall low degeneracy. Therefore, the associated mode capacity and expressivity are poor, as compared to the GSP embedding even at small system sizes.

Finally, we visualize scaling for spectral properties for the GSP-embedding and contrast it to the rotational embedding. Choosing the spectral degree as a metric, in Fig.~\ref{spectrum}(d) we plot as a blue curve the $D_{\Omega_{\mathrm{R}}} \propto N^2$ scaling for the tower-type feature map. We compare this with the GSP-based feature map, remembering that frequencies scale with $T \propto \Delta_{\min}^{-2}$, and thus the annealing time has a strong system size dependence. At the very least the minimal annealing gap drops as $\Delta_{\min} \propto N^{-1}$ for simple instances of the ground state preparation \cite{AlbashLidarRMP2018}. We show this as an orange curve in Fig.~\ref{spectrum}(d), where $D_{\Omega} \propto N^3$ (and stress that the degeneracy is largely superior to the rotation-based embedding). The second scenario is the exponentially minimal annealing gap closing as $\Delta_{\min} \propto 2^{-N/2}$, the typical scaling for hard instances \cite{AlbashLidarRMP2018}. This case is depicted as a dark red curve in Fig.~\ref{spectrum}(d), and further highlights the difference of associated QML models.

\subsection*{Coefficients of the Ising ground state feature map}

We have seen that the quantum model based on ground states produced by adiabatic evolution \eqref{eq:f_model} can be written as a truncated Fourier series, $f_{\theta}(x) = \sum_{\omega \in \Omega} c_\omega \exp[-i\omega x]$, with the degree of the spectrum bounded by the model eigenvalues, total evolution time and number of Trotter steps. But what about the coefficients $c_\omega$? Their size is determined by the $\hat{w}_{\ell}^{(m)}$ operators, which depend on both the $x$-dependent and $x$-independent operators, as well as the measurement operator $\hat{\mathcal{M}}_{\theta}$. The magnitude of the coefficients gives us information about which frequency modes are dominant in a particular QML model, and their variance defines model expressivity.

\begin{figure}[ht]
\includegraphics[width=1.0\linewidth]{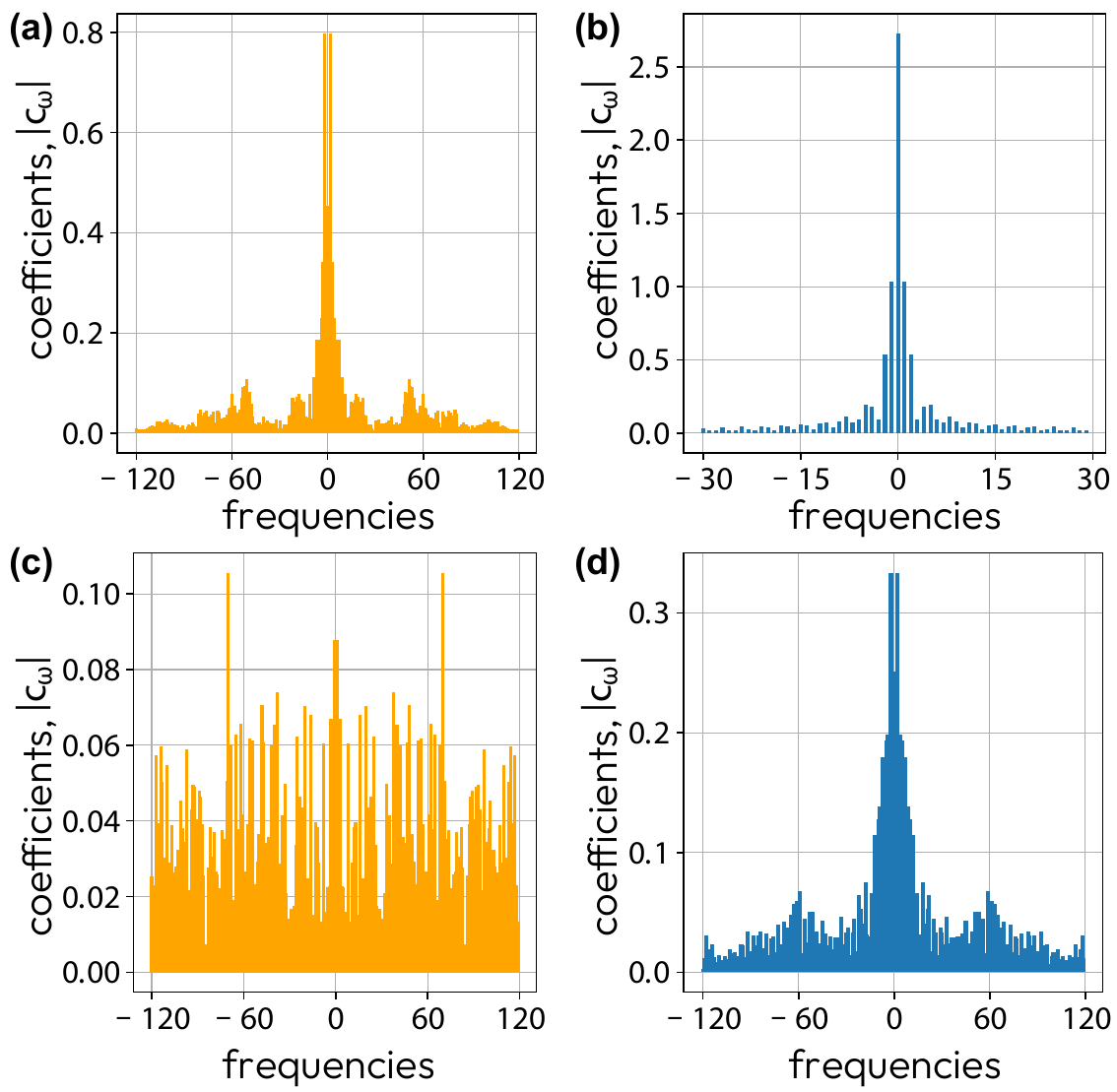}
\caption{Absolute value of the coefficients of the Fourier model from the Trotterized adiabatic ground state preparation. We consider GSP protocols for: $M=100$, $T=100$ \textbf{(a)}; $M=1000$, $T=100$ \textbf{(b)};  $M=100$, $T=1000$ \textbf{(c)}; and $M=1000$, $T=1000$ \textbf{(d)}.}
\label{coefficients}
\end{figure}

To investigate the expressivity of ground state-based QML models, we set $\hat{\mathcal{M}}_{\theta}$ as the magnetization operator, $\hat{H}_{\mathrm{Z}}$, with no variational circuit in between the adiabatic ground state embedding and the cost operator. This ensures that our model $f_{\theta}(x)$ is an approximation to the phase boundary observed in Fig.~\ref{adiabatic}(a). We know that the model has an integer frequency spectrum because the $x$ dependence is embedded into $\hat{H}_{\mathrm{Z}}$, therefore we can accurately calculate the corresponding coefficients using a fast Fourier transform. We show these coefficients in Fig.~\ref{coefficients}.

Just as the fidelity of the adiabatically-prepared ground states are dependant on both the total evolution time $T$ and the number of Trotter steps $M$, the same is also true for the size and spread of the coefficients. When both $T$ and $M$ are small, the quantum model gives an inaccurate approximation to the phase boundary, and the coefficients are more spread across all frequencies. An increase in $M$ improves the ground state fidelity significantly, and this manifests itself in an increase in the ratio between low and high-frequency coefficients. We can explain this by looking at the gap spectrum in Fig. \ref{spectrum}: there is a large degeneracy in the low frequency modes, so these low frequencies have many more terms contributing to their coefficients. An increased $M$ enhances this effect, causing the spectrum to become mode peaked. 

We also note that even with high $M$, if we increase $T$ so that the time step $\Delta t$ is too large, the ground state approximation becomes poor and the coefficients become more spread out again, as seen in Fig.~\ref{coefficients}(d), where $\Delta t = 1000/1000 = 1$.


\section{Discussion}

Once we have formally introduced the ground state-based embedding for quantum machine learning models, and analyzed their properties, we are left with a question: where do we go from here? Ideally, we require QML workflows that are motivated by preparing non-trivial input states (see Cerezo et al. \cite{Cerezo2023CSIM}) while making sure they are advantageous from the learning point of view. The presented analysis and results show that unitary processes that include large number of re-uploadings, but have non-variational circuits in between, can effectively mimic GSP-type feature maps. This can be a direction to explore from the QML modelling point of view (capacity and expressivity), as well as classical simulability (testing if a quantum state follows the area law of entanglement). Also, we foresee that effective Hamiltonians for data embeddings can be designed, such that the basis functions are smooth even away from the large Trotter number limit (for instance, by playing with commutativity of terms that comprise $\hat{\mathcal{H}}(x)$).

Another point of discussion is the universality of our analysis for the GSP embeddings. A careful reader might wonder if the described scaling for the spectrum is a consequence of choosing the adiabatic preparation, and its potentially non-optimal $N$-dependence on the annealing time and the number of steps. We show that this is not the case by presenting in the Appendix an alternative analysis, which leads to the same conclusion about GSP-based QML models being high-frequency models with massive degeneracies and very structured coefficients.


\section{Conclusions}

In this study we have presented and formalized a quantum data embedding based on the ground state preparation protocols. We have represented a problem as a digital version of the adiabatic ground state preparation, which allows us to connect GSP-based feature maps with rotation-based maps and Fourier-type QML models. We analyzed properties of the ground state-based quantum machine learning models, revealing that ground state-based feature maps can be seen as models with large capacity, where the frequency gap spectrum grows at least as a high-degree polynomial in the number of qubits $N$. For complex Hamiltonians this scaling of the degree and size of the spectrum becomes exponential. We observed that the spectrum is largely degenerate, and frequencies are tied with coefficients for the emerging basis that are highly structured. This leads to highly specific models with limited expressivity, where formally large model capacity is effectively tamed. Given the trade-offs between expressivity and trainability \cite{Holmes2022}, we expect GSP-type maps to be valuable for increasing trainability. Our results point a way for QML modelling with provably non-simulable input states and improved generalization.


\section*{Acknowledgment}

The authors thank Vincent E. Elfving, Annie E. Paine, and Ivan Rungger for useful discussions. We acknowledge the funding from UK EPSRC (award EP/Y005090/1) and thank NATO SPS project MYP.G5860 for supporting this work.

\appendix*

\section{Alternative analysis of GSP feature maps}
\label{Appendix}

In the main text we have based the ground state quantum feature map analysis on the adiabatic state preparation. One of the main conclusions highlights the large degree and massive degeneracies of the associated spectrum. Let us show that these conclusions hold, specifically by presenting an alternative analysis.

Let us assume that we have full access to the feature state vector $|\psi_{\mathrm{G}}(x)\rangle$. We can construct a (rather artificial) operator $\hat{\mathcal{U}}^{\mathrm{FF}}_{\theta}(x) = \exp[-i \theta \hat{\mathcal{G}}(x)]$ such that for selected $\theta^*$ and a tailored generator $\hat{\mathcal{G}}(x)$ we directly map the input state to the ground state, $|\psi_{\oslash}\rangle \mapsto |\psi_{\mathrm{G}}(x)\rangle$. For this, we extend the register with one ancillary qubit, such that $|\psi_{\oslash}\rangle \rightarrow |\psi_{\oslash}\rangle \otimes |0\rangle \equiv |\psi_{\oslash}^{\mathrm{(e)}}\rangle$ and $|\psi_{\mathrm{G}}(x)\rangle \rightarrow |\psi_{\mathrm{G}}(x)\rangle \otimes |1\rangle \rangle \equiv |\psi_{\mathrm{G}}^{\mathrm{(e)}}(x)\rangle$. In this case we have the initial state and the target state being orthogonal, $\langle \psi_{\oslash}^{\mathrm{(e)}}|\psi_{\mathrm{G}}^{\mathrm{(e)}}(x)\rangle = 0$. We form the generator $\hat{\mathcal{G}}(x) = |\psi_{\mathrm{G}}^{\mathrm{(e)}}(x)\rangle \langle \psi_{\oslash}^{\mathrm{(e)}}| + | \psi_{\oslash}^{\mathrm{(e)}}\rangle \langle \psi_{\mathrm{G}}^{\mathrm{(e)}}(x)|$ such that this enables a rotation in the relevant state space. Note that the squared generator is a projector, $\hat{\mathcal{G}}^2(x) \equiv \hat{P}_{\mathrm{G}}(x)$, $\hat{\mathcal{G}}^4(x) = \hat{P}_{\mathrm{G}}^2(x) = \hat{P}_{\mathrm{G}}(x)$, and that due to state orthogonality $\hat{\mathcal{G}}(x) \cdot \hat{P}_{\mathrm{G}}(x) = \hat{\mathcal{G}}(x)$. With this we can expand $\hat{\mathcal{U}}^{\mathrm{FF}}_{\theta}(x)$ as a Taylor series and collect the terms as
\begin{align}
\label{eq:UFF}
\hat{\mathcal{U}}^{\mathrm{FF}}_{\theta}(x) = \left[\mathbb{1} - \hat{P}_{\mathrm{G}}(x)\right] + \cos(\theta) \hat{P}_{\mathrm{G}}(x) - i \sin(\theta) \hat{\mathcal{G}}(x).
\end{align}
Setting $\theta \rightarrow \theta^* = \pi/2$ we have the unitary state preparation schedule $|\psi_{\mathrm{G}}^{\mathrm{(e)}}(x)\rangle = \hat{\mathcal{U}}^{\mathrm{FF}}_{\pi/2}(x) |\psi_{\oslash}^{\mathrm{(e)}}\rangle$ (up to a global phase), and the ancilla qubit can be disregarded.

What does this artificial GSP give us in terms of the spectral analysis of the feature map? We observe that the fast-forwarded GSP operator can be rewritten as a generator $\hat{\mathcal{G}}(x) = \sum_{k=1}^{4^N} \varphi_k(x) \hat{\mathcal{P}}_k$ decomposed into Pauli words $\hat{\mathcal{P}}_k$ with $x$-dependent functions $\varphi_k(x)$. Here the number of possible words goes up to $4^N$, and as we decompose operators with support on few states, there is a high likelihood that exponentially many words are indeed involved (as they resemble Behemoth operators discussed in \cite{Khaymovich2019}). Next, we can group Pauli words into commuting groups $\mathcal{S} = \big\{ \{\hat{\mathcal{P}}_{k \in \Gamma}\} \big\}_{\Gamma}$, but the cardinality of $\mathcal{S}$ for a generic problem will still grow rapidly with $N$ \cite{Gokhale2019,Shlosberg2023}. Therefore, representing the fast-forwarded GSP operator still requires Trotterization with the number of steps $M_\mathcal{S}$ dependent on $|\mathcal{S}|$ and the norm of operators in each group $\{\hat{\mathcal{P}}_{k \in \Gamma}\}$, yielding
\begin{align}
\label{eq:U_FF_Pauli}
\hat{\mathcal{U}}^{\mathrm{FF}}_{\pi/2}(x) \approx \prod_{m=1}^{M_\mathcal{S}} \left[ e^{-i \frac{\pi}{2 M_\mathcal{S}} \varphi_1(x) \hat{\mathcal{P}}_1} \cdot \ldots \cdot e^{-i \frac{\pi}{2 M_\mathcal{S}} \varphi_{4^N}(x) \hat{\mathcal{P}}_{4^N}} \right].
\end{align}
Each Pauli word contains only $\pm 1$ eigenvalues, and the degree of the associated spectrum scales as $\mathcal{O}(4^N)$, unless the generator is highly structured. Also, given that $\varphi_k(x)$ have nonlinear dependence (which is usually the case, for instance looking at the Ising model basis), the underlying basis is non-Fourier. We conclude that one cannot represent quantum models based on ground state-based feature maps as simple Fourier models \cite{Landman2022random}. 



%

\end{document}